\magnification=1200
\vsize=22 true cm
\hsize=16 true cm
\baselineskip= 0.6 true cm

\line{\bf  LYCEN 9721   \hfill May 1997}

\bigskip
\parindent=0 true cm
\pageno=1

\def\hor{\hskip 1.0 true cm}
\def\gsim{\mathrel{\rlap{\lower4pt\hbox{\hskip1pt$\sim$}}
    \raise1pt\hbox{$>$}}}         

\vglue 2. true cm
\centerline{\bf LOCAL NUCLEAR SLOPE and CURVATURE}
\centerline{\bf in HIGH ENERGY $pp$ and $\bar pp$ ELASTIC SCATTERING}
\bigskip
\centerline{\bf P. Desgrolard$^{(1)}$, J. Kontros$^{(2)}$,
A.I. Lengyel$^{(2)}$, E.S. Martynov$^{(3)}$.}

\bigskip

\centerline{$^{(1)}$ Institut de Physique Nucl\'eaire de Lyon, IN2P3-CNRS}
\centerline{et Universit\'e Claude Bernard, F 69622 Villeurbanne Cedex, France}
\centerline{(e-mail~: p.desgrolard@ipnl.in2p3.fr)}

\medskip
\centerline{$^{(2)}$ Institute of Electron Physics, 294016 Universitetska, 21
Uzhgorod, Ukraine}
\centerline{(e-mail~: iep@iep.uzhgorod.ua)}

\medskip
\centerline{$^{(3)}$ Bogolyubov Institute for Theoretical Physics}
\centerline{National Academy of Sciences of Ukaine}
\centerline{Metrologichna st., 14b, 252143 Kiev, Ukraine}
\centerline{(e-mail~: martynov@gluk.apc.org)}
\bigskip
\bigskip

\bigskip
{\bf Abstract}
\medskip

\hor
The local nuclear slope $B(s,t) = {d \over d t}
\left(\ln {d\sigma_n (s,t)\over dt}\right)$
is reconstructed from the experimental angular distributions
with a procedure that uses
overlapping $t$-bins , for an energy that ranges from the ISR
to the $S\bar ppS$ and the Tevatron.
Predictions of several models of ($p,p$) and
($\bar p,p$) elastic scattering at high energy are tested in $B(s,t)$
at small $|t|$.
Only a model with two-components
Pomeron and Odderon gives a satisfactory agreement with
the (non fitted) slope data, in particular for the evolution of $B(s,t)$ with
$s$ as a function of $t$ in $\bar pp$ scattering.
This model predicts a similar behavior for $pp$ and $\bar pp$ scattering at
small $|t|$.
A detailed confirmation for $pp$ collisions would be expected from RHIC.

\hor
The extreme sensitivity of the local nuclear curvature
$C(s,t) = {1\over 2} {d \over d t}\left (B(s,t)\right)$
with the choice for a Pomeron model is emphazised.
The present model predicts a change of sign for  $C(s, t= 0)$
when $\sqrt s \gsim 4 $ TeV. The ideal place to search for an eventual
confirmation of this prediction would be LHC.
\bigskip
\bigskip

            \vfill\eject

{\bf 1. Introduction}
\bigskip

\hor
In  ($p,p$) and ($\bar p,p$) elastic scattering, the local nuclear
slope parameter (for brevity "slope"), defined  from the nuclear part
of the differential cross-section  ${d\sigma_n (s,t)\over dt} $ as~:
$$B(s,t)\ = \ { d \over d t}\left(\ln {d\sigma_n (s,t)\over dt}\right)
,\eqno(1)$$
is a sensitive tool for investigating the fine structure of the first
diffraction cone of the angular distribution at small squarred
4-momentum transfer $|t|$.  It is known that this structure presents
two characteristic patterns (at a fixed energy $\sqrt{s}$)~:
{\it (i)} a "break" [1-3], {\it i.e.} a change $\Delta B$ $\sim 2$ GeV$^{-2}$
around $|t|\sim 0.1$ GeV$^2$ at all energies (not yet confirmed at the
Fermilab Tevatron)

{\it (ii)} localized fluctuations [4-6] (also called "oscillations")
over a smooth background, which may reach a 10\% ratio [6]
on $B$-values, while they are limited to 2\% on
$d\sigma\over dt$-values.
Whether this latter pattern, sometimes controversial, signals new interesting
and unexpected physics awaits experimental confirmation [4].

\hor
In addition to the slope, the local nuclear curvature parameter (simply
called "curvature"), related to the second derivative of the angular
distribution by~:
$$C(s,t) =  {1\over 2}
\left({d \over\ d t} (B(s,t)\right)               \eqno(2)$$
is a meaningful characteristic of the slope.
Actually, we shall consider this quantity essentially at $t= 0$ and write
$C_0(s)= C(s,t= 0)$. This curvature at $t= 0$
has been considered for a long time by Block and Cahn [7a] as a
"sensitive indicator of the transition to asymptoptia", going from
positive values at the ISR energies, to near zero at the Tevatron
energy.  They have expected for any model (in particular their model
{\it \`a la}  Chou-Yang [7b]), which approaches  the sharp black disk
limit) that $C_0(s)$ should become very  negative at higher energies.
This has not been confirmed by computations of slopes in [8, 9] and of
course by experiments.

\hor
The sudden decrease of $C_0(s)$ when $\sqrt s$ increases near the
Tevatron has been also reported in data analyses (see for example the
work of Pumplin [10], where it is found that the zero value is
consistent with the experiment).  This is certainly a remarkable and
interesting result, but unfortunately extracting $C(s,t)$ from
available angular distributions data is not currently done (probably
because large errors are induced) and direct precise measurements are
still lacking.

\bigskip

\hor
The purpose of this note is to check how a detailed reproduction of the
slope data (usually non fitted) allows to discriminate among some theoretical
models and to predict the slope in the kinematical range of future
experimental projects PP2PP [11] at RHIC and TOTEM [12] at LHC.
However, we shall not be concerned in the present paper with the reproduction
of the above mentioned fluctuations of the slope over its background.

\hor
Finally, we intend to test the above mentioned prediction concerning the sign
change of the curvature at $t= 0$
when the energy exceeds $\sim 2$ Tev in both cases $pp$ and $\bar pp$.
\bigskip
\bigskip
{\bf 2. Model for the scattering amplitude}
\bigskip

\hor Among the available models for
($p,p$) and ($\bar p,p$) elastic scattering, we choose the model of
Jenkovszky and collaborators [13] essentially because of its simplicity
(the calculations
are made at the "Born level") and of its evolution with the time towards
aversion giving a high quality fit to all the data .
Since the original works where the Pomeron is a double-pole (called "dipole"
for brevity) in the complex angular momentum $J$ plane at $J= \alpha (t)$,
several generalizations have appeared (see [14-17] and references therein).

\hor
In its simplest form, the dipole model for the Pomeron, complemented later
by  the Odderon and two Reggeons, describes succesfully the ISR data [14].
In particular, it reproduces well the "dip-bump" structure of the $pp$
differential cross-section. The inclusion of the Reggeons is required to
reproduce the lowest energy data (at $t= 0$) and that of the Odderon to fill
the shoulder in the $\bar pp$ differential cross-section.

\hor
In more sophisticated versions, on the one hand the dipole concept has
been generalized
towards a multipole Pomeron [15] (in fact limited to a triple-pole,
a "tripole" for brevity) and on the other hand,  nonlinear
trajectories [16] have been used for the Pomeron.
However, unhappily, the attempts to improve the quality of the fit and
to extend the kinematical $s$- and $t$-range of its application have
resulted, as usual, in a complication of the model and consequently in
an increasing number of free parameters.
Nevertheless, one has to go beyond the dipole approximation,
if one wants a really good fit for all existing data.
It has been shown [17] that one can variously combine the points of view
of {\it (i)}
Donnachie and Landshoff [18] for the "supercritical" Pomeron, of {\it (ii)}
Gauron {\it et al.} [19] for the "asymptotic" Pomeron and
Odderon respecting the asymptotic theorems
and of {\it (iii)} Jenkovszky and collaborators [13,14] for the Pomeron
(Odderon) amplitude as a double or triple pole.

\hor
We recall that in the investigated versions of the above model, the
total cross-sections rise at very high energy with the Pomeron
contribution as $s^\epsilon \ln {s/s_0}$ [14] or as $ \ln^2 {s/s_0} $
[17] for the dipole versions and as $s^\epsilon \ln^2{s/s_0}$ [15,17]
for the tripole version ($\epsilon =  \alpha(0)-1$, with the Pomeron
intercept $\alpha(0)$).  These behaviors may be compared with the
traditional logarithmic rises in $\ln {s/s_0}$ and $\ln^2 {s/s_0}$ and
with the power-like rise in $s^\epsilon$ due to Donnachie and Landshoff
[20], which has now become a standard reference [21], supported also
by consideration on WA91 glueball [22].
The adequacy of a given rise relies on the values of the parameters,
which in turn depend on the chosen amplitude and on the fitted energy
and transfer ranges. In our (conservative) opinion, a family of
possible rises will persist until very high energies will be
experimentally investigated (total cross-sections at LHC rather than
more precise measurements of structure functions at HERA) where,
ultimately, the Froissart-Martin bound will have to be obeyed.

\hor
Useful formulae for the amplitude are summarized in the Appendix.

\hor
We investigated five successive versions of the model {\it \'a la}
Jenkovszky {\it et al.} (see Table 1).
For the unpublished version (II), we changed only the linear trajectory used
 in (I) into a logarithmic one and performed the calculations as in [14].
The version (III) and (IV) are studied in [15] and [16] respectively.
For the published version (V) [17], we choose the option where the so-called
(see [19]) "minimal"
amplitude  has been added to the complete tripole amplitude with a
linear trajectory.  In Table 2 we compare, as a possible measure of the
quality of the fit, the $\chi^2/$d.o.f. for the various versions.
For the purpose of comparison, in each case the
parameters have been refitted over the same set of ($\sim$ 1000) data
covering the $\sqrt s $-range from 4 to 1800 GeV for total
cross-sections and $\rho$-value at $t= 0$ and the $\sqrt s $-range from
23 to 630 GeV, the $|t|$-range from 0 to 14 GeV$^2$ extended to
differential cross-sections. Note that the Tevatron angular
distribution was not included in the fit.

\bigskip
\bigskip

{\bf 3. Procedure for obtaining the slope from the data.
Results of the analysis.}
\bigskip

\hor
The slopes $B(s,t)$ can be reconstructed from the
available  differential cross-sections data. For each investigated
energy, they are represented by a set of "experimental" slopes $b_i$
obtained on small ranges of transfer $|t_i|$ (bins) with a reasonable
number of points
($\sim $ 10). Contrary to the procedure traditional in the studies of
the break as by Schiz {\it et al.} [23], the  bins are shifted by one
or more channel so that they overlap (for more details, see [24], where
preliminary results are also presented).  Within the $i^{\rm th}$ bin
used for fitting the differential cross-sections $({d\sigma\over
dt})_i$, we set neglecting the Coulomb contribution
$$({d\sigma\over dt})_i\ \simeq\ a'_i e^{b_it},    \eqno(3) $$
where $a'_i$ and $b_i$ are free parameters. As a consequence (and this
is the interest of using overlapping bins), we obtained a number of
"experimental" values $ b_i\pm\Delta b_i$ close to the original number
of true available experimental points for ${d\sigma\over dt}$.
The error bars $\Delta b_i$
represent the fitting uncertainty.

\hor
In the Coulomb interference region, (3) must be replaced by
$$({d\sigma\over dt})_i\ = \ \left|a_i e^{b_i t/2}(\rho_i + i) +
F_c\right|^2,  \eqno(4)$$
where the first term represents the nuclear contribution ($\rho_i$ is the
ratio of the real to the imaginary part of the forward nuclear amplitude)
and the second term $F_c$ is the standard Coulomb amplitude, which
can be calculated with a good approximation according to the procedure by
West  and Yennie [25]. This way, one
extra parameter ($\rho_i$) is added, but one may fix it to its
experimental value at $t= 0$ to keep meaningful the fitting procedure
with a limited number of points.

\hor
Using this procedure we reconstruct the $pp$ and $\bar pp$
"experimental slope" $ b_i\pm\Delta b_i$ for available ISR and
colliders energies and for $-t$ up to 3 GeV$^2$. A selection of our
results plotted versus $-t$ is shown in Fig.1 (see also Table 3,
where indications are given to compare with previous analysis [10,23]
and experimental data [26]). We observe the following features of the
structure of the slope~:

{\it (i)} the collapse of the slope at $|t|\sim 1$ GeV$^2$
corresponding to  the dip (for $pp$) or shoulder (for $\bar pp$) in the
angular distribution is of course found out at all energies, except at
the Tevatron, where the measurements are limited by a too small
$|t|$-upper bound

{\it (ii)} at ISR energies and for $pp$ and $\bar pp$,
when $|t|$ increases, the experimental slope behaves
from zero up to $|t|\sim 0.5$ GeV$^2$, as a slowly decreasing sequence of
local values distributed along a smooth curve with oscillations around it

{\it (iii)} in the $\bar pp$ case, we can see by eye directly on the
plots of the experimental slope that the curvature $C_0(s)$ would be
decreasing with $\sqrt s$, going from positive values to almost zero~:
in other words, for very high energies, the mean curve $B(s,t)$
slightly decreases for small $|t|$ at the CERN Collider and is almost
flat at the Tevatron.
The limited energy range of the available data prevents us from establishing
on experimental ground the same tendency for $pp$ scattering.

\hor We do not intend to present an extensive analysis of the curvature
$C(s,t)$ versus $t$ at given energies as in [10], mainly because
extracting  the curvature from the data merely introduces new errors.
Generalizing the above method and extrapolating $C$ at some finite
$t$ to $t= 0$ would not be reliable.  However, due the important
character of the sign change of $C_0(s)$, we quote in Table 4 previous
results [10] for a few energies.  We simply note that the definition of
the  curvature (2) implies at small $-t$ a parametrization of the
differential cross-section of the type
$$({d\sigma\over dt})\ \simeq\ a e^{B_0 t+C_0 t^2},  \eqno(5) $$
where $B_0 $ and $C_0$
are the local slope and curvature at $t= 0$.

\bigskip
\bigskip
{\bf 4. Theoretical results and discussion}
\bigskip

\hor
The slope and the curvature have been calculated from
the nuclear scattering amplitude $A= A(s,t)$ given by the five versions
quoted above via (1,2) or explicitly
$$ B(s,t)= {2\over |A|^2} {\Re{\hbox{e}}}\left(A^*{\partial A\over
\partial t}\right)  \eqno(6)$$
$$ C(s,t)= {1\over |A|^2} {\Re{\hbox{e}}}\left(A^*{\partial^2 A\over\partial
t^2} +\left|{\partial A\over\partial t}\right|^2 \right)
-{1\over 2} B^2(s,t) .    \eqno(7)$$
We first turn our attention to the slope $B(s,t)$.

\hor
The simplest case "dipole with linear trajectory" (I) gives,
for all investigated energies (up to 1.8 TeV), constant slopes for
$|t|$ up to 0.5 GeV$^2$ for both $pp$ and $\bar pp$,
(shifted upwards with increasing $\sqrt s$~). Changing the linear trajectory
into a non-linear as in (II) modifies only the curvature which takes
roughly the same positive values independent of the energy.  For the
"tripole with linear or logarithmic trajectory" (III and IV), the
situation is almost the same~: there is no change of the shape of
$B(s,t)$, plotted versus $t$  for $|t|$ up to 0.5 GeV$^2$;
the energy introduces only an overall shift of the curves.
These remarks exclude the pure dipole and tripole models for
reproducing, and consequently for extrapolating, $B$-values. For this
reason, we do not report the corresponding results.

\hor
On the contrary, the version "V" of the model with two components
(complete tripole + minimal Pomeron) reproduces not only the slope at
$t= -0.02$ Gev$^2$ versus the energy as shown in [17] but all the
available slope data.  Representative plots are selected in Fig.1,
where they are compared to the results of our analysis (see preceeding
section).  In particular, the agreement is very good up to the dip for
$pp$ and includes the shoulder for $\bar pp$.  The very delicate dip
mechanism is better reproduced when the energy increases and the
dip's height increases.

\hor
An overall comparison with other published results is difficult,
because calculations have been performed in partial and different
transfer-energy ranges. However, we show in
Table 3 a selection of previous computed results [7a,8b,9a,27]
and results of data analysis [10,23] for $B(s,t= 0)$ compared to ours.
One notices again the excellent agreement found between our theoretical and
experimental results.

\hor
The most impressive result of our calculations shown in Fig.1 concerns
the behavior of the slope versus $|t|$, which is found to be energy
dependent and in agreement with the experimental features of our data
analysis mentioned above, thus confirming our confidence in the validity
of the model presented in [17].
Of course, we do not pretend to reproduce the complicated experimental
oscillations of the slope with respect to  smooth curves~:
it is known that this implies a special parametrization for low $|t|$
(see for example the "toy model" of [6b]).
Neither, we shall not discussed their origin and their period as in [6b].

\hor
We remark that, not suprisingly, it is the version of the multiple
Pomeron model giving the best overall fit (including high $|t|$
differential cross-sections), which also reproduces best the low
$|t|$ (non fitted) slopes.  The question might be why it works so well.
A tentative explanation may be related on the role of the Odderon.

{\it (i)} Real difficulties arise when fitting
all existing data (from $|t|= 0$ to highest $|t|$, up to
$\sqrt s = 1800$ GeV) and a success in doing so [17] probably means
that the choice of the amplitude, which includes the Odderon, is not
devoided of physical sense.

{\it (ii)} It is well known that the Odderon contribution affects most
strongly the large $|t|$-data (see for example [19,17]), although it
has recently be shown that $t= 0$-data alone may be able to decide about
the existence of a relevant Odderon  [28]. Therefore, if in view of this
theoretical result (still waiting for
an experimental confirmation) we suspect a correlation between the successes
of a model (with an Odderon) at large and small $|t|$ at the same time.

{\it (iii)} The influence of the Odderon also appears in comparing the
different structures of the slope in the dip region at the ISR energies for
$pp$ and $\bar pp$~: they reflect the existence of a real
dip in the angular distributions for $pp$ and of a shoulder for $\bar pp$
which we have mentioned earlier [17].

\hor
In addition, for $\bar p p$ scattering, the model "V" gives a sign change
of the curvature at $t= 0$ with energy~: {\it i.e} a decreasing slope
at very small $|t|$ for energies below the Tevatron becomes
progressively a rising slope (corresponding to $C_0<0$) for higher
energies. An examination "by eye" of Fig.2 does not contradict this result,
which is also in qualitative agreement with the trend exhibited by our
slope analysis of the data at $\sqrt s \le 1.8$ GeV, confirming
predictions of a model {\it \`a la } Chou and Yang calculated by Block
and Cahn [7a].  Searching the origin of this sign change, we found that
it is due (see the Appendix) to the "minimal" contribution
($A_{min}^+(s,t)$, rising asymptotically with energy
$\propto s\ln s$) which has been added to the complete tripole Pomeron
($A_{I}^+(s,t)$, rising asymptotically as $\propto s^{1+\Delta}\ln^2 s$)
and which, at high energies, gives a negative $C_0-$value,
though, strickly speaking, $C_0\rightarrow 0$,
due to the dominating tripole Pomeron contribution at $s\rightarrow
\infty$.  However, this occurs numerically at an energy which can
compared with Planck's one.
To clarify we plotted in Fig.3a, our calculated curvature $C_0(s)$
versus the energy $\sqrt{s}$. After presenting a maximum at low energy,
the curve exhibits that the "onset of asymptotia", corresponding to the
vanishing $C_0(s_{as})= 0$, would be at $$\sqrt{s_{as}}\sim 4 \hbox{
TeV}.$$ The calculated value of $C_0$ becomes substantially negative at
the LHC energy.

\hor
To get more confidence in our theoretical predictions for
$C_0$, we compare in Table 4 to previous calculations [7a,27] and data
analysis [10]. A model dependence is found even for models giving
a good agreement with the angular distribution data. This is the case of
[29], where the chosen parametrization does not allow to show a
dependence of the curvature with the energy. This is an illustration
of the interest of considering the slope and curvature at small $t$ when
one is concerned with very fine characteristics of the angular distribution.
Seeking further for the model dependence of the curvature at $t= 0$
versus  the energy, we find out very strongly different behaviors.
Aside from the main model under consideration (version "V", see above)
two extremal cases of asymptotic behaviors $C_0(s)\rightarrow
\pm\infty$ when $s\rightarrow \infty$ are also found. One of them,
$C_0(s)\rightarrow -\infty$, encountered in [7] and [19], is discussed
above. The opposite case, $C_0(s)\rightarrow +\infty$ is predicted in
the so-called dipole Pomeron model with a nonlinear trajectory (see
[30] for details). This model is constructed for small $|t|$ (first
diffraction cone) and it leads to high quality description of $pp$
and $\bar pp$ differential cross-sections in the domain $\sqrt{s}\ge
9$ GeV and $|t|\le 0.5 $GeV$^2$.
We show the behavior of $C_0(s)$ calculated in that case in Fig.3b , which
is to be compared to the result of Fig.3a.  It
is necessary to remark that, within this model framework, the
zero curvature is seen only at $t\neq 0$ at the Tevatron energy.
In this  model the value of the local curvature $C(s,t)$  for
$|t| > $ 0.1 GeV$^2$  is small and
closed to zero in the domain of the first diffraction cone.
Therefore a large value of $C_0(s)$ does not contradict
available experimental data.

\hor
Furthermore, the comparaisons of both the local values at $t\ne 0$
for  the slope $B(s,t)$ and the curvature $C(s,t)$ show in Table 5 a
less rapid decrease of our theoretical results with $|t|$ with respect
to those of the analysis by Pumplin [10].

\smallskip

\hor
Then, a natural question arises~:  what is the the energy
dependence of the slope, in particular concerning the change in sign of
the curvature at $t= 0$ for $pp$ scattering at energies beyond the
ISR where the experimental data are now lacking~?

\hor
From the experimental point of view, the RHIC [11] and LHC projects [12]
are the ideal machines to answer this question. RHIC is
expected to provide $pp$ data between $\sqrt s = 60$ GeV and
$\sqrt s = 500$ GeV in the
$|t|$-range from 0.005 to 6 GeV$^2$, and will thus be a very useful
complement to the available $\bar pp$ data~;
in particular, the projected slope measurements in
the nuclear Coulomb interference region will allow to follow
the evolution of
$B(s,t)$ plotted versus $t$ as the energy increases.
LHC [12] will produce $pp$ collisions at the energies
never obtained so far ($\sqrt s = 10$ TeV to 14 TeV),
covering the kinematical range $|t|\sim 0.01$ to 8-10 GeV$^2$ and will
provide an unambigous answer about the sign of the
curvature at small $|t|$.

\hor
From the theoretical point of view, our extrapolation of the calculated
slope  for $pp$ is shown in
Fig.4 for various energies of these projects for the best version ("V")
of our model. The same remarks as in the $\bar pp$ case apply
for the energy dependence and for the change in sign of the
curvature at $t= 0$ (which presents a similar behavior in both $pp$ and
$\bar pp$, see also Fig.3).
Finally, the model predicts an abrupt fall of the $pp$
slope at $|t|$-values decreasing to 0.6 GeV$^2$ when $\sqrt s$
exceeds 10 TeV, corresponding to a pronounced dip.
Whether these predictions will pass the experimental test of future
measurements is, of course, the crucial question.

\bigskip
\bigskip

{\bf 5. Conclusions}
\bigskip

\hor
We have shown that a detailed calculation of the non fitted nuclear
slope data allows to discriminate among a family of versions of a
theoretical scattering model.  We have analysed a model giving an
excellent fit to $\bar pp$ and $pp$ data to check what this gives for
this one piece of data which has not been used in the fit, {\it i.e.}
the local nuclear slopes reconstructed from the angular distributions.
The result of the best version is excellent for $pp$ and $\bar pp$
for which data exist, allowing us to make predictions for $pp$ which
should be tested in future machines~:  in particular a change of sign
at $\sim 4$ TeV for the local curvature at $t= 0$.

\hor
Of course, a prerequisite is the availability of very precise
experimental data at small $t$, allowing to extract the local slope and
curvature and extrapolate to $t= 0$.  They appear as  very sensitive
quantities, to be used as a test to select among realistic models of
elastic hadron scattering.

\bigskip
\bigskip
{\bf Acknowledgements}
\medskip

We are indebted to M. Giffon for his continuous support.
We thank J.O. Pumplin for interesting correspondance.

                              \vfill\eject

\centerline{\bf APPENDIX}

\bigskip

\hor
For the sake of clarity and of
being self-contained, we give a brief summary of the formulae
used for calculating the nuclear scattering amplitude in the version
giving the best fit (see [17] for more details).

\hor
The four other versions
may be viewed as particular cases of the one presented here, following the
indications of the text, summarized in Table 1; their performances are
compared in Table 2.

\medskip
\hor
The most general amplitudes we used in that paper, fitted to reproduce
$\bar pp$ and $pp$ elastic
scattering data on total cross-section, real to imaginary
part of forward amplitudes and angular distributions
($\sigma_t $, $\rho$ and ${ {d \sigma}\over {dt} }$, respectively)
are decomposed into a crossing even
and a crossing odd contribution ($A^+$ and $A^-$) as
$$A_{pp}(s,t) \, = \, A^+ - A^-, \quad   A_{\bar p p}(s,t) \, = \,
A^+ + A^-.  \eqno(A1)$$
Each contribution $A^\pm$ is conveniently splitted into two components
$A^{\pm}_{I,II}$ according to
$$A^{\pm} =  A^{\pm}_{I} + A^{\pm}_{II} \, .  \eqno(A2)$$
Firstly, $A^{\pm}_{I}$ will be constructed as a tripole (or a
dipole in simpler cases) for the Pomeron ($P$) and the Odderon ($O$)
corrected by two secondary Reggeons ($f$, $\omega$) . More precisely, we
write
$$A^{+}_{I} =  A_P + A_f ,\quad A^{-}_{I} =  A_O + A_\omega.  \eqno(A3)$$
Secondly, $A^{\pm}_{II}$  will be constructed according
to a Gauron {\it et al.} [19] (GLN) prescription.

We use the normalisation
$\sigma_{t0t} =  { {4 \pi}\over s}\ \Im\hbox{m} A (s,t= 0)$.
\bigskip

\hor
To construct the first component of the amplitude, we choose the
standard $f$- and $\omega$-Reggeons, parameterized as:
$$A_f (s,t)= a_f \, e^{\alpha _f (t)\ln \tilde s } ,\quad A_\omega (s,t)= i\,
a_\omega \, e^{\alpha _\omega (t)\ln \tilde s }, \eqno(A4)$$
where we  have defined, as usual $\tilde s =   {s\over{s_0}}\, e^{-{i\pi\over
2}}.$ In what follows, we choose the scale parameter~: $s_0 =  1$
GeV$^2$ and the Reggeon trajectories as in [17]~: $\alpha _f (t) =
0.69 + 0.84 \, t$ and
$\alpha _\omega (t) =  0.47 + 0.93 \, t$, with $t$ in GeV$^2$.

\medskip
\hor
For the Pomeron, we use the {\it complete} tripole
({\it i.e.} we include also the monopole and the dipole contributions).
The same construction will apply {\it mutatis mutandis} to the
Odderon. Then, we write the corresponding Pomeron amplitude (a label
$P$ for Pomeron and $O$ for Odderon necessary to distinguish the two
amplitudes in (A3) is understood) as
$$ A(s,t)= \sum^3_{n= 1}\; c^{(n)}\,A_n\,(s,t)\; \; ,  \eqno\hbox{(A5)}$$
where each individual contribution $A_n(s,t)$ is given by
$$
A_n(s,t)=
{d^{n-1}\over d\alpha^{n-1}}\,
\big [e^{\alpha\ell n\tilde s}\,
G(\alpha)\big ]\; \; ,
\eqno\hbox{(A6)}
$$
and
$$\alpha =  \alpha (t) =  1+\epsilon +\alpha't \,  \eqno(A7) $$
is the linear trajectory with an intercept $1+\epsilon$.
For the residue $G(\alpha)$ we define ("tripole ansatz")
$$
{d^3\,G(\alpha)\over d\alpha^3}=
b^3\,e^{b(\alpha-1)}\; \; .
\eqno\hbox{(A8)}
$$
To obtain the amplitude, we have to integrate three times to get,
subsequently
$G''(\alpha)$, $G'(\alpha)$ and $G(\alpha)$. Lastly, we sum the amplitudes
corresponding to the three multipoles. This yields the
following amplitude for the complete tripole
$$
A (s,t)= e^{\alpha (t)\ell n\tilde s}\; [\; \; ] \; \; ,
\eqno\hbox{(A9)}
$$
where
$$
\eqalign{
[\; \; ]  & =
e^{b(\alpha(t)-1)}\;
\big (c^{(1)}+c^{(2)}\,(b+\ell n\tilde s)+
c^{(3)}\,(b+\ell n\tilde s)^2\big )\cr
& + D_1(t)\,(c^{(1)}+c^{(2)}\,\ell n\tilde s+
c^{(3)}\,\ell n^2 \tilde s)+
D_2(t)\,(c^{(2)}+2c^{(3)}\,\ell n\tilde s)\ + d^{(3)} c^{(3)} \; .\cr
}\eqno\hbox{(A10)}
$$
According to the previous developments, the functions $D_i(t)$ are now
expressed in terms of the trajectories as
$$\eqalignno{
D_1(t)= &d^{(1)}+d^{(2)}\,\alpha(t)+{1\over 2}d^{(3)}\,\alpha^2(t)\; ,
& \hbox{(A11)}\cr \cr
D_2(t)= &d^{(2)}+d^{(3)}\,\alpha(t)\; , & \hbox{(A12)}\cr
}$$
where the constants $d^{(i)}$ ($i= 1,2,3$) are parameters coming from the
successive integrations performed to get $G$.

\hor
The Odderon contribution to $A^-_I$ is constructed analogously.
The modification with respect to the Pomeron contribution to $A^+_I$
is a multiplicative factor
$i$ in the amplitude and of course a change of the parameters.

\medskip

\hor
As for the (additional) second component of the amplitude,
following the same kind of prescription as in [19],
one can derive the following expressions for the Pomeron
(which we call, by analogy {\it minimal} since it corresponds to
a $\ell n s$ asymptotic growth in the total cross section) and denote by
$A_{min}^+$
$$ A_{II}^+ =
A_{min}^+=   -\tilde s\left[P_1\ell n\tilde s{\sin(R \, \tilde\tau)
\over R \, \tilde\tau} \, \beta^{(1)}(t)
       +P_2 \cos(R \, \tilde\tau) \, \beta^{(2)}(t)\right],
\eqno(A13)$$
and for the Odderon (due to its small contribution) in that version the
second part is simply cancelled $$A_{II}^- = 0.  \eqno(A14)$$
In the above equation, $\tilde\tau =  \sqrt{-{t\over t_0}}\ln\tilde s$,
with $t_0$ fixed at 1 GeV$^2$, and contrary to GLN [7] we choose all
the functions
$$\beta^{(i)}(t)=  (1-{t\over \eta^{(i)}})^{-4}\, ;i= 1,2 \ ,\eqno(A15)$$
so as to satisfy the perturbative QCD requirements.

\hor
The various parameters of this "complete tripole plus minimal Pomeron"
version are listed in Table 1 of [17].  As already said,  the above
formulae can be reduced (see Table 1)  to construct the amplitudes for
any of the other versions we have consider in that paper, by choosing
appropriately the "coupling constants", the  various integration
constants entering in  the residue $G$  and the form of the trajectories,
either linear (A7) or logarithmic [16]

$$\alpha =  \alpha (t) =  \alpha_0+\alpha_1 t -\alpha_2
\ln{\left( 1-\alpha_3 t\right)} \ .   \eqno(A16)$$

              \vfill\eject

\centerline{\bf References}
\bigskip

\item{ [ 1]} G. Barbierini {\it et al.}~: Phys. Lett. {\bf B} (1972) 663.
\smallskip

\item{ [ 2]} M. Bozzo {\it et al.}~: Phys. Lett. {\bf B} (1984) 3851.
\smallskip

\item{ [ 3]} UA4/2 collaboration, C. Augier {\it et al.}~: Phys. Lett.
{\bf B 316} (1993) 448.
\smallskip

\item{ [ 4]} J.N.J. White~: Nucl. Phys. {\bf B51} (1973) 23.
\smallskip

\item{ [ 5]} C. Bourrely, J. Soffer, T.T. Wu~: Phys. Lett. {\bf B 315}
(1993) 195.
\smallskip

\item{ [ 6]} -a)
S. Barshay, P.Heiliger~:  Z. Phys. {\bf C 64} (1994) 675.

-b) P. Gauron, B. Nicolescu, O.V. Selyugin~: Phys. Lett. {\bf 397} (1997)
305.
\smallskip

\item{ [ 7]} -a)
M.M. Block, A.N. Cahn~: Phys. Lett. {\bf B 149} (1984) 245.

-b) T.T. Chou and C.N. Yang~: Phys. Rev {\bf 170} (1968) 1591.
\smallskip

\item{ [ 8]} -a) R. Glauber, J. Velasco~: Phys. Lett. {\bf B 147} (1984)
380;

-b) {\it ibid.}~:
Proceedings of the 2$^{\rm th}$ International Conference on Elastic and
Diffractive Scattering, (2$^{\rm th}$ "Blois workshop", October 1987),
Rockfeller University, New York. K. Goulianos, editor, p 219.
\smallskip

\item{ [ 9]} -a) C. Bourrely, J. Soffer, T.T. Wu~: Nucl. Phys. {\bf B247}
(1984) 15;

-b) {\it ibid.}~: Z. Phys. {\bf C 37} (1988) 369; Phys. Lett. {\bf B
252} (1990) 287.
\smallskip

\item{[10]} J. Pumplin~: Phys. Lett. {\bf B 276} (1992) 517.
\smallskip

\item{[11]} W. Guryn  in "Frontiers in Strong Interactions",
VII$^{\rm th}$ Blois Workshop on Elastic and
Diffractive Scattering, Chateau de Blois, France - June 1995,
Edited by P. Chiappetta, M. Haguenauer, J. Tran Thanh Van, Editions
Fronti\`eres (1996), p 419.
\smallskip

\item{[12]} M. Buenerd in "Frontiers in Strong Interactions",
VII$^{\rm th}$ Blois Workshop on Elastic and
Diffractive Scattering, Chateau de Blois, France - June 1995,
Edited by P. Chiappetta, M. Haguenauer, J. Tran Thanh Van, Editions
Fronti\`eres (1996), p 437 and references therein.
\smallskip

\item {[13]} L.L. Jenkovszky~: Fortsch. Phys. {\bf 34} (1986) 702.

L.L. Jenkovszky, A.N. Shelkovenko, B.V. Struminsky~: Z. Phys. {\bf C 36}
(1987) 495 and references therein.
\smallskip

\item {[14]} P. Desgrolard, M. Giffon, L.L. Jenkovszky~: Z. Phys.
{\bf C 55} (1992) 637.
\smallskip

\item {[15]} P. Desgrolard, M. Giffon, L.L. Jenkovszky~: Phys. At. Nucl.
{\bf 56} (October 1993) 1429.
\smallskip

\item{[16]} P. Desgrolard, L.L. Jenkovszky, A.I. Lengyel~:
{\it in} "Strong interactions at long distances", L.L. Jenkovszky
editor, Hadronic Press, Palm Harbor (1995), p.235.

\smallskip
\item{[17]} P. Desgrolard, M. Giffon, E. Predazzi~: Z. Phys. {\bf C 63}
(1994) 241.

\smallskip
\item{[18]} A. Donnachie, P.V. Landshoff~: Nucl. Phys. {\bf B231} (1984)
189; {\bf B244} (1984) 322; {\bf B267} (1986) 406.

\smallskip
\item{[19]} P. Gauron, B. Nicolescu, E. Leader~: Nucl. Phys. {\bf B 299}
(1988) 640; Phys. Lett. {\bf B 238} (1990) 406.

\smallskip
\item{[20]} A. Donnachie, P.V. Landshoff~: Phys. Lett. {\bf B296}
(1992) 227.

\smallskip
\item{[21]} Review of Particle Properties~: Phys. Rev. {D50} (1995) 1335.

\smallskip
\item{[22]} WA91 Collaboration (S. Abatzis {\it et al.})~: Phys. Lett.
{\bf B324} (1994) 509.

J.R. Cudell, K. Kang, S.K. Kim~: BROWN-HET-1016 (Jan. 96),
hep-ph/9601336.

\smallskip
\item{ [23]} A. Schiz {\it et al.}~: Phys. Rev. {\bf D 24} (1981) 26.

\smallskip
\item{[24]} J. Kontros, A.I. Lengyel~:
{\it in} "Strong interactions at long distances", L.L. Jenkovszky
editor, Hadronic Press, Palm Harbor (1995), p.67.

P. Desgrolard, J. Kontros, A.I. Lengyel~: {\it " Multipole Pomeron
models and prediction of the local nuclear slope"},
Institut de Physique nucl\'eaire Universit\'e Claude Bernard \hfill\break
Lyon I, LYCEN 9603 (March 1996).

\smallskip
\item{[25]}
G.B. West, D.R. Yennie~: Phys. Rev {\bf 172} (1968) 1413.

\smallskip
\item{[26]} -a) UA4 Collaboration~: Phys. Lett. {\bf B 344} (1995) 451.

-b) CDF Collaboration~: Phys.Rev. {\bf D 50} (1994) 5518.

-c) E710 Collaboration~: Phys. Rev. Lett. {\bf 68} (1992) 2433.

\smallskip
\item{[27]} E. Gotsman, E.M. Levin, U. Maor~: Z. Phys. {\bf C 57}
(1993) 677.

\smallskip
\item{[28]} M. Giffon, E. Predazzi, A. Samokhin~: Phys. Lett.{\bf B 375}
(1996)  315.

\smallskip
\item{[29]} P.V. Landshoff~: DAMPT 96/48 (May 1996), hep-ph/9605383.

\smallskip
\item{[30]}
P. Desgrolard, A.I. Lengyel, E.S. Martynov~: {\it " Nonlinearities and
Pomeron nonfactorizability in conventional diffraction"},
Institut de Physique nucl\'eaire Universit\'e Claude Bernard Lyon I,
LYCEN 9640 (November 1996). hep-ph/9704404, {\it to be published in}
Nuov. Cim. {\bf A}.

                                  \vfill\eject

\def\init{\tabskip 0pt\offinterlineskip}
\def\crr{\cr\noalign{\hrule}}

$$\vbox{\init\halign to 13 cm{
\strut#&\vrule#\tabskip= 1em plus 2em&
\hfil$#$\hfil&
\vrule#&
\hfil$#$\hfil&
\vrule#&
\hfil$#$\hfil&
\vrule#\tabskip 0pt\crr
&&{\hbox{\bf Version}}&&\hbox {\bf Trajectory eq.} &&
\hbox{\bf Amplitude} &\crr
&&{\hbox{   I}  }&&(A7)&& c^{(1)}=  c^{(3)}=  d^{(2)}=  d^{(3)}= A_{II}^\pm =  0 &\cr
&&{\hbox{  II}  }&& (A16)&&\hbox{ idem}  &\cr
&&{\hbox{ III}  }&& (A7)&&  A_{II}^\pm =  0  &\cr
&&{\hbox{  IV}  }&& (A16)&& \hbox{ idem}  &\cr
&&{\hbox{   V}  }&&(A7)&& \hbox {general formulae}  &\crr
}}$$
\centerline {\bf Table 1}
Explaining how to deduce the various versions of the model from the
general formalism given in the Appendix.

\bigskip

$$\vbox{\init\halign to 13 cm{
\strut#&\vrule#\tabskip= 1em plus 2em&
\hfil$#$\hfil&
\vrule#&
\hfil$#$\hfil&
\vrule#&
\hfil$#$\hfil&
\vrule#&
\hfil$#$\hfil&
\vrule#\tabskip 0pt\crr
&&  &&{\hbox{\bf Pomeron\ Model}}&& {\bf Ref.}&&{\chi^2 /\hbox {d.o.f.}}
&\crr
&&{\hbox{I}}  &&{\hbox{ Dipole  with linear trajectory}  }&&[14]&
                    &    15.&\cr
&&{\hbox{II}} &&{\hbox{ Dipole  with non linear trajectory}  }&&
\hbox{unpub.}&
&10. &\cr
&&{\hbox{III}}&&{\hbox{ Tripole  with linear trajectory}  }&&[15]&
                    &    4.5&\cr
&&{\hbox{IV}}&&{\hbox{ Tripole  with non linear trajectory}  }&&[16]&
&3.7 &\cr
&&{\hbox{V}}&&{\hbox{ Tripole  +  "minimal"}  }&&[17]&
&2.8 &\crr
}}$$

\centerline {\bf Table 2 }
 $\chi^2$ results for the five successive versions of the model listed
 in Table
I. The parameters
for each version have been refitted over the same set of ($\sim 1000$)
data (see the text).

                   \vfill\eject

\bigskip

$$\vbox{\init\halign to 13 cm{
\strut#&\vrule#\tabskip= 1em plus 2em&
\hfil$#$\hfil&
\vrule$\,$\vrule#&
\hfil$#$\hfil&
\vrule#&
\hfil$#$\hfil&
\vrule#\tabskip 0pt\crr
&&\hbox{\bf Energy (GeV)}&&{\bf B}(s,0) \hbox{ (GeV}^{-2}) \hbox{
(previous)}
&&\quad\hbox{(present)}&\crr
&&19.4 \ \qquad pp   && \hbox{anal.}&&\hbox {anal.}&\cr
&&  && 11.74\ [23] &&      12.5\pm 1.5      &\cr
&&    && 12.44\pm 0.04\ [10]  &&           &\cr
&&    &&  && \hbox {th.}&\cr
&&  &&   && 12.3 &\crr
&& \qquad \qquad\bar pp    && \hbox{exp.}&&    &\cr
&&541\ \ \ \qquad   && 15.5\pm 0.1\ \hbox{[26a]}   &&  &\cr
&&546\ \ \ \qquad && 15.28\pm 0.58\ \hbox{[26b]} &&    &\cr
&&  && \hbox{anal.}&&\hbox {anal.}&\cr
&&   && 16.82 \pm 0.22\ [10] &&    15.6\pm 0.3  &\cr
&&    && \hbox{th.}&&\hbox {th.}&\cr
&&    &&16.7\pm 0.7\ [7a] &&  15.2            &\cr
&&    &&15.6\ \ [9a]         &&    &\cr
&& &&18.0\ [8b]        &&     &\crr
&&1800 \qquad \bar pp   && \hbox{exp.}&&        &\cr
&&  && 16.98\pm 0.25\ \hbox{[26b]}&&   &\cr
&&  && 16.99\pm 0.47\ \hbox{[26c]}&&  &\cr
&&  && \hbox{anal.}&&\hbox {anal.}&\cr
&&  &&\sim 16.4\ [10]   &&  16.20 \pm .23     &\cr
&&    && \hbox{th.}&&\hbox {th.}&\cr
&&    &&19.5\pm 1.4\ [7a] && 16.3   &\crr
&&40000\qquad pp    && \hbox{th.}&&\hbox {th.}&\cr
&&    &&  28.3\pm 3.8\ [7a]  &&  19.8 & \cr
&&    &&  20.7\ [9a]  &&   &\cr
&&    &&  19.-22. [27]  &&   &\crr
}}$$

\centerline {\bf Table 3 }

Local nuclear slope at $t= 0$..
The "previous" values are from experiments (exp. [26]) or from
analysis of experimental data (anal. [10,23]) or
from theoretical calculations (th. [7a,8b,9a,27]). The "present"
ones are those issued  from the present analysis (anal.) and those
calculated (th.) with the version "V" of the model (see the text).

\vfill\eject

$$\vbox{\init\halign to 13 cm{
\strut#&\vrule#\tabskip= 1em plus 2em&
\hfil$#$\hfil&
\vrule$\,$\vrule#&
\hfil$#$\hfil&
\vrule#&
\hfil$#$\hfil&
\vrule#\tabskip 0pt\crr
&&\hbox{\bf Energy (GeV)}&&{\bf C}(s,0) \hbox{ (GeV}^{-4}) \hbox{
(previous)}
&&\quad\hbox{(present)}                           &\crr
&&19.4\ \qquad pp     &&\hbox{anal.}  &&          &\cr
&&       &&7.72\ [10]&&                           &\cr
&&   &&  \hbox {th.}    &&\hbox {th.}             &\cr
&&   &&  \sim 4.\ [27]       && 7.3               &\crr
&&546 \ \qquad \bar pp   && \hbox{anal.}&&        &\cr
&&           &&13.65\ [10] &&                     &\cr
&&    && \hbox{th.}&&\hbox {th.}                  &\cr
&&    && 3.9\ [7a]&& \ 5.2                        &\cr
&&           &&\sim 0.5\ [27] &&                  &\crr
&&1800 \qquad \bar pp   && \hbox{anal.}&&         &\cr
&&  &&\ \ \sim 0.0\ [10]\ &&                      &\cr
&&    && \hbox{th.}&&\hbox {th.}                  &\cr
&&    && -0.03\ [7a]\ && 2.7                      &\cr
&&           &&\sim -2.\ [27] &&                  &\crr
&&40000 \quad pp   && \hbox{th.}&&\hbox {th.}     &\cr
&&   && -25.2\ [7a]\ && -7.0                      &\cr
&&           &&\sim -13.\ [27] &&                 &\crr
}}$$

\centerline {\bf Table 4}

Local nuclear curvature at $t= 0$.
The "previous" values are from the analysis of experimental data by
Pumplin (anal. [10]) or from theoretical calculations by Block and Cahn
(th. [7a]) and extracted from Fig.4 of Gotsman {\it et al.} [27].
The "present" ones are those calculated (th.) with the version "V"
of the model (see the text).

\vfill\eject

$$\vbox{\init\halign to 13 cm{
\strut#&\vrule#\tabskip= 1em plus 2em&
\hfil$#$\hfil&
\vrule$\,$\vrule#&
\hfil$#$\hfil&
\vrule#&
\hfil$#$\hfil&
\vrule$\,$\vrule#&
\hfil$#$\hfil&
\vrule#&
\hfil$#$\hfil&
\vrule#\tabskip 0pt\crr
&&-t\hbox{\bf (GeV}^2)&&{\bf B} \hbox{ (GeV}^{-2}) [10]
&&\ \hbox{present}
&&{\bf C} \hbox{ (GeV}^{-4}) [10]
&&\ \hbox{present}
&\crr
&&0.00     &&12.44  &&12.27&&7.72&&7.33        &\cr
&&0.02     &&12.15  &&12.01&&7.04&&6.73        &\cr
&&0.04     &&11.88  &&11.77&&6.44&&6.19        &\cr
&&0.06     &&11.63  &&11.56&&5.90&&5.72        &\cr
&&0.08     &&11.41  &&11.36&&5.42&&5.29        &\cr
&&0.10     &&11.20  &&11.18&&4.98&&4.92        &\cr
&&0.20     &&10.38  &&10.49&&3.31&&3.59        &\cr
&&0.30     && 9.84  &&10.03&&2.17&&2.82        &\cr
&&0.40     && 9.50  && 9.71&&1.28&&2.32        &\cr
&&0.50     && 9.32  && 9.50&&0.48&&1.93        &\crr
}}$$

\centerline {\bf Table 5}

Comparaison of our present results calculated with the version
"V" of the model with those of the analysis from Pumplin [10] for the
$pp$ slope $B(s,t)$ and curvature $C(s,t)$ at $\sqrt{s}= 19.4 $ GeV, as
functions of $t$.



                     \vfill\eject

\centerline{ \bf Figures Captions}
\bigskip

{\bf Figure 1.}
Results of our analysis of local nuclear slope $b_i\pm\Delta b_i$
reconstructed from experimental $\bar pp$ and $pp$
angular distributions with the procedure of overlapping bins (see Ref.[24]
and the text, Sect.2).
The solid lines are slopes $B(s_0,t)$ calculated with the two components
version "V" of the model (complete tripole + minimal Pomeron).
The slope is plotted versus $-t$ (logarithmic scale) for
selected energies ($\sqrt s_0$).
\bigskip

{\bf Figure 2.}
Calculated slope $B(s_0,t)$
plotted versus $-t$ (linear scale) for $\bar pp$ elastic scattering.
The version "V" of the model is used for $\sqrt s_0$= 53 GeV, 546 GeV,
1.8 TeV, 10 TeV.
See also Fig.1, where some results of our data analysis are shown
(no data exist at energies higher than the Tevatron energy).
\bigskip

{\bf Figure 3.}
Calculated curvature at $t= 0$, $C_0(s)$ plotted versus the energy
$\sqrt{s}$
for $\bar pp$ (solid line) and $pp$ (dashed line) elastic scattering.

(a) The version "V" of the model is used.

(b) The model of [30] is used.

\bigskip

{\bf Figure 4.}  Calculated slope $B(s_0,t)$
plotted versus $-t$, with the version "V" of the model,
in the $pp$ case from the ISR energy  $\sqrt s_0 = 60$ GeV, extrapolated
to the 500 GeV of the PP2PP project (at RHIC) [11]
and to $\sqrt s_0 = 10$ TeV, 14 TeV of the TOTEM project (at LHC) [12].



\vfill\eject

\input epsf
\centerline{\bf Figure 1}
\epsfxsize= 15.2 true cm
\centerline{\epsfbox{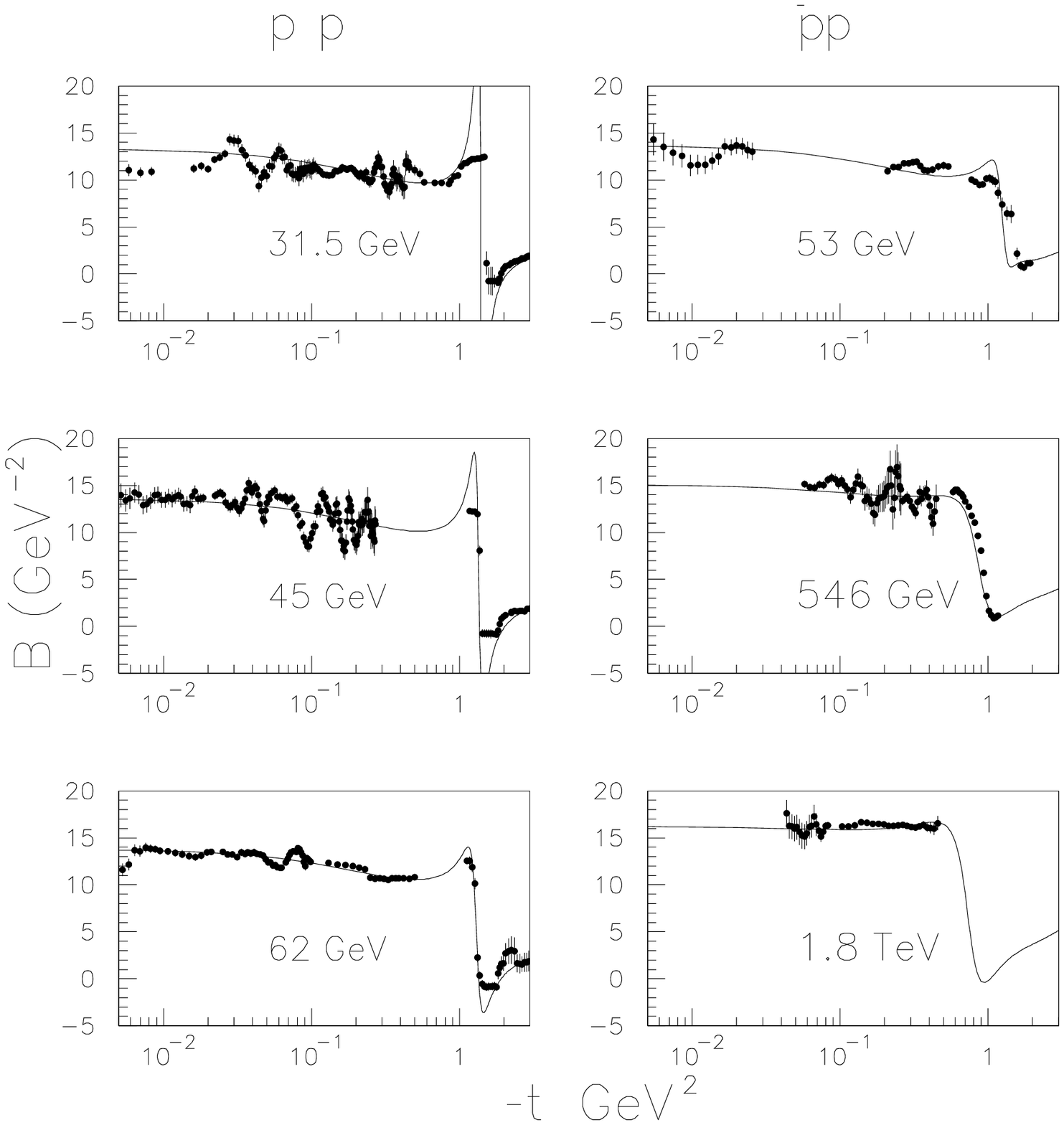}}
\centerline{\bf Figure 2}
\epsfxsize= 15.2 true cm
\centerline{\epsfbox{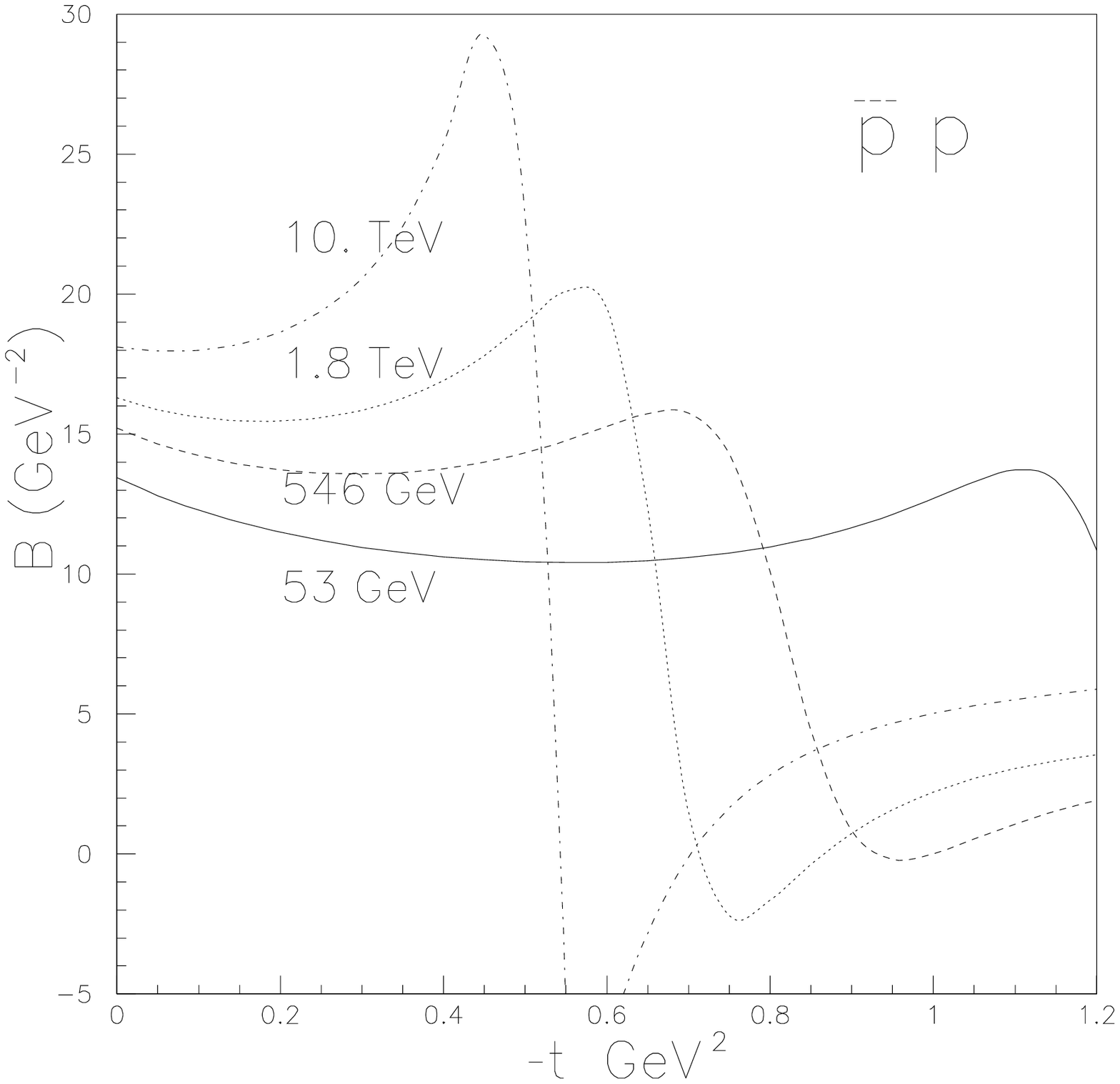}}
\centerline{\bf Figure 3a}
\epsfxsize= 15.2 true cm
\centerline{\epsfbox{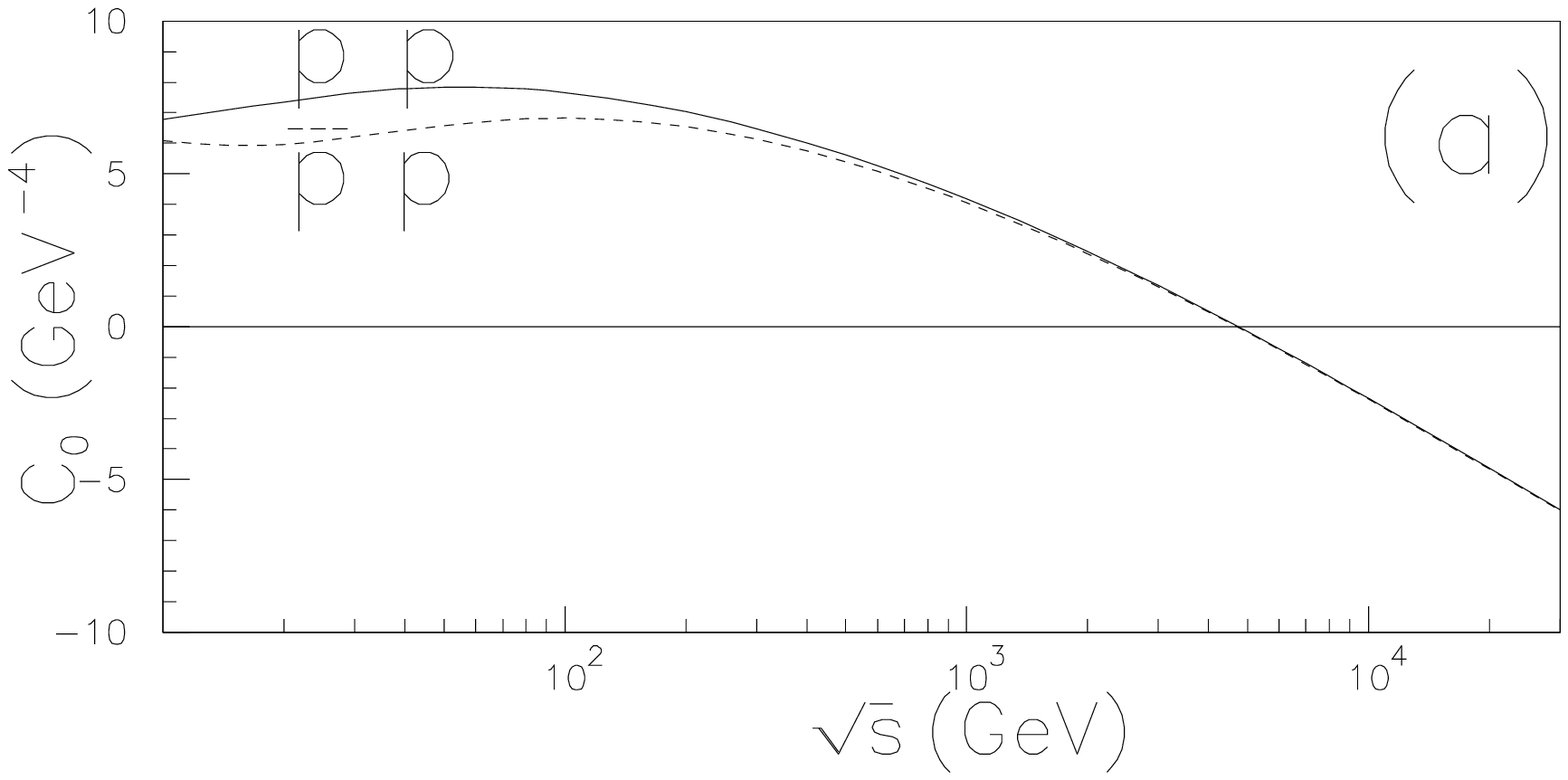}}
\centerline{\bf Figure 3b}
\epsfxsize= 15.2 true cm
\centerline{\epsfbox{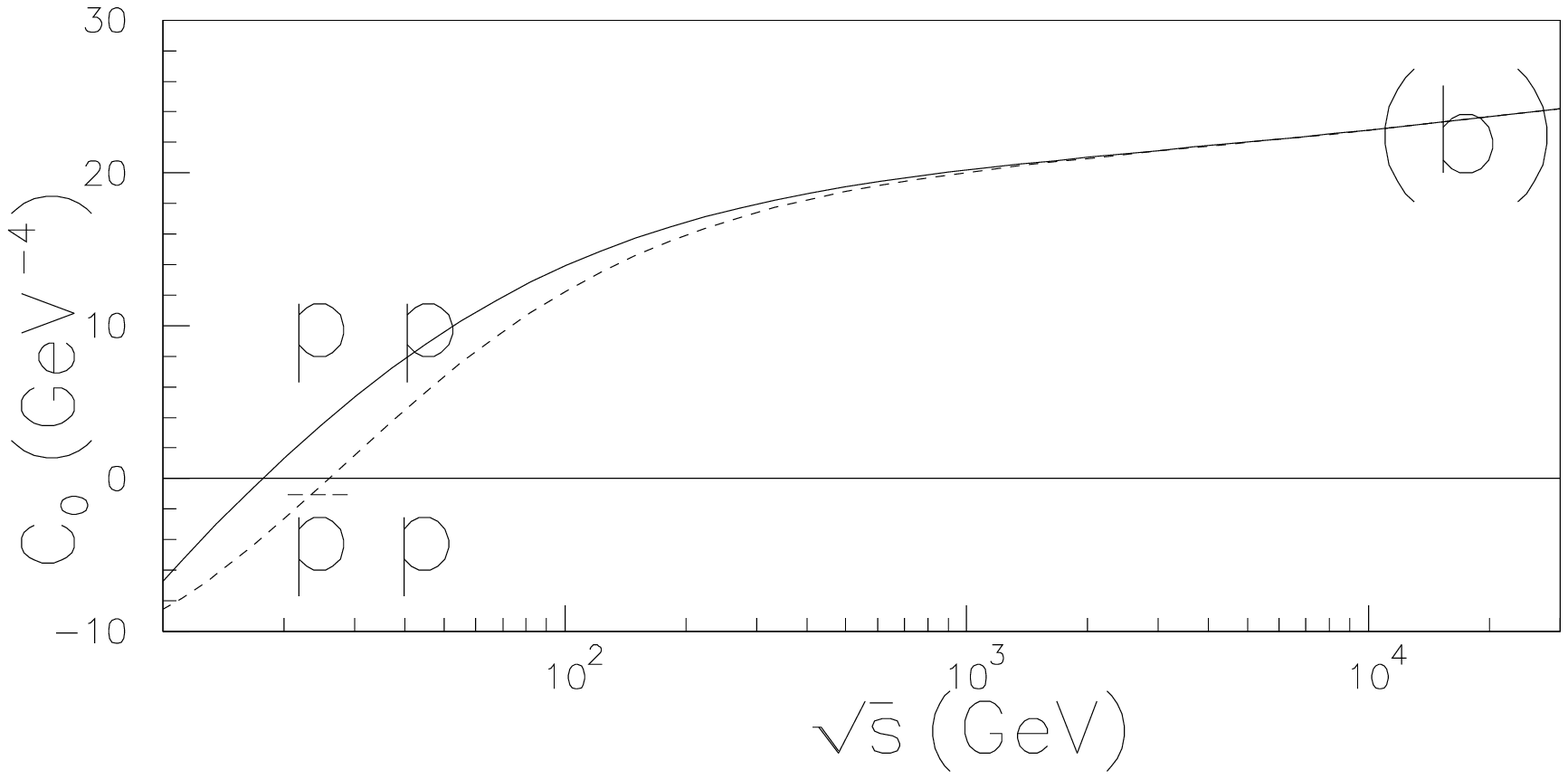}}
\centerline{\bf Figure 4}
\epsfxsize= 15.2 true cm
\centerline{\epsfbox{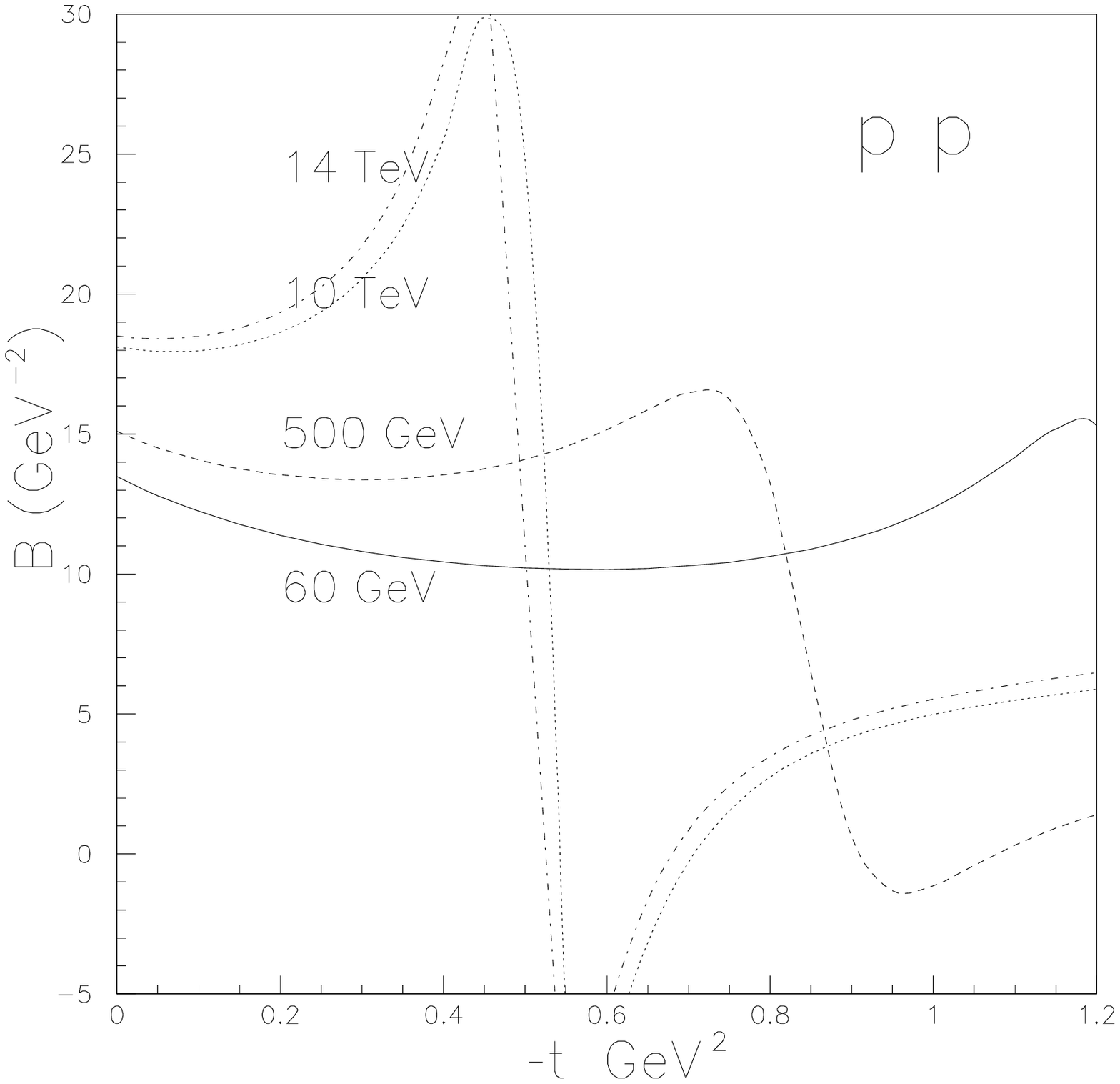}}

     \bye